\documentclass[sn-mathphys,Numbered]{sn-jnl}

\usepackage{graphicx}%
\usepackage{multirow}%
\usepackage{amsmath,amssymb,amsfonts}%
\usepackage{amsthm}%
\usepackage{mathrsfs}%
\usepackage[title]{appendix}%
\usepackage{textcomp}%
\usepackage{manyfoot}%
\usepackage{booktabs}%
\usepackage{algorithm}%
\usepackage{algorithmicx}%
\usepackage{algpseudocode}%
\usepackage{listings}%
\usepackage[skip=20pt]{caption}
\usepackage[table,xcdraw]{xcolor}
\usepackage[normalem]{ulem}
\useunder{\uline}{\ul}{}
\usepackage[graphicx]{realboxes}
\usepackage{framed}

\usepackage{hyperref}%
\usepackage{xcolor}%

\theoremstyle{thmstyleone}%
\theoremstyle{thmstyletwo}%

\theoremstyle{thmstylethree}%

\begin{document}

\title[Article Title]{Leveraging genomic deep learning models for the prediction of non-coding variant effects}

\author[1]{\fnm{Pooja} \sur{Kathail}}
\equalcont{These authors contributed equally to this work.}

\author[2]{\fnm{Ayesha} \sur{Bajwa}}
\equalcont{These authors contributed equally to this work.}

\author*[1,2,3]{\fnm{Nilah M.} \sur{Ioannidis}}\email{nilah@berkeley.edu}

\affil[1]{\orgdiv{Center for Computational Biology}, \orgname{University of California Berkeley}, \orgaddress{\city{Berkeley, CA}, \country{USA}}}

\affil[2]{\orgdiv{Department of Electrical Engineering and Computer Sciences}, \orgname{University of California Berkeley}, \orgaddress{\city{Berkeley, CA}, \country{USA}}}

\affil[3]{\orgname{Chan Zuckerberg Biohub}, \orgaddress{\city{San Francisco, CA}, \country{USA}}}

\abstract{Characterizing non-coding variant function remains an important challenge in human genetics. Genomic deep learning models have emerged as a promising approach to enable \textit{in silico} prediction of variant effects. These include supervised sequence-to-activity models, which predict molecular phenotypes such as genome-wide chromatin states or gene expression levels directly from DNA sequence, and self-supervised genomic language models. Here, we review progress in leveraging these models for non-coding variant effect prediction. We describe practical considerations for making such predictions and categorize the types of ground truth data used to evaluate variant effect predictions, providing insight into the settings in which current models are most useful. Our Review highlights key considerations for practitioners and opportunities for improvement in model development and evaluation.}

\maketitle


\section*{Main}\label{intro}

Interpreting the functional consequences of non-coding variants remains a challenge in human genetics, with important applications to rare disease, cancer, and complex traits. Genome-wide association studies (GWAS) identify relatively common and largely non-coding variants associated with complex traits and diseases, but interpreting GWAS results requires disentangling causal variants from non-causal variants in linkage disequilibrium (LD) and identifying the genes, cell types, and tissues affected \cite{welter2014nhgri, cano2020gwas}. Rare, \textit{de novo}, and somatic non-coding variants can have large functional impact \cite{Zappala2016} and play important roles in rare disease and cancer.
Machine learning methods have been successfully applied to predict the impact of protein-coding variants \cite{adzhubei2010method, ioannidis2016revel, cheng2023accurate, pejaver2022calibration}, and similar early approaches for non-coding variants, including CADD (Combined Annotation Dependent Depletion) \cite{Rentzsch2018} and k-mer based models \cite{Ghandi2014, lee2015method}, have also shown promise \cite{ritchie2014functional, shihab2015integrative, Huang2017, rojano2019regulatory, ellingford2022recommendations}. For non-coding variants, functional genomics data—such as measurements of gene expression or chromatin accessibility—have become a rich source of information on active genomic regions in disease-relevant cell types. These datasets also provide training data for modeling non-coding sequences with genomic deep learning models, which are neural networks that aim to learn the function of DNA sequences. These sequence-based models have the advantage of making predictions based on a learned, functional understanding of genomic sequence and can handle any input sequence containing arbitrary lengths or combinations of variants, allowing them to broadly predict the functional impact of non-coding variation.

While the design, training, and interpretation of genomic deep learning models have been reviewed in detail elsewhere \cite{Angermuller2016, Zou2018, Eraslan2019, Wang2024review, sasse2024unlocking, sokolova2024deep, capitanchik2024computational, Consens2025, Benegas2025LMreview, Novakovsky2022}, here we focus specifically on the use of such models for variant effect prediction, which is distinct from their promising and complementary role in dissecting reference genome regulatory grammar. A detailed review of progress and best practices in this specific application area is needed, since many studies have begun to apply these models to variant effect prediction in practice. For example, several studies have used sequence model-based variant effect predictions to prioritize likely causal variants within trait-associated loci \cite{Zhou2018, Schwessinger2020, Hudaiberdiev2023}, which can contain complex combinations of causal variants \cite{Abell2022}, and incorporating model predictions as priors in functionally-informed fine-mapping \cite{kichaev2014integrating, schaid2018genome} is one promising direction \cite{Wang2021}. Variant effect predictions from genomic deep learning models have also been used to prioritize \textit{de novo} mutations in individuals with Autism Spectrum Disorder \cite{Zhou2019, Jaganathan2019, Trevino2021, Linder2022, marderstein2025mapping} and congenital heart disease \cite{Richter2020, Ameen2022}, with experimental validation of the prioritized variants in some cases \cite{Zhou2019, Jaganathan2019, Ameen2022}. Finally, model variant effect predictions, along with model interpretability techniques \cite{Shrikumar2017, Lundberg2017, simonyan2014inputgradient, Novakovsky2022}, have been used for mechanistic interpretation of disease-relevant variants \cite{Zhou2018, Chen2022, Yang2022, Alipanahi2015, Kim2021, celaj2023bigrna, Park2021}, such as predicting the cell types and transcription factor motifs most likely affected \cite{Trevino2021, Ameen2022, Wang2022, loeb2024variants}.

Here we review the application of genomic deep learning models to non-coding variant effect prediction, including progress in evaluating performance across a range of biological scales, from variant effects on molecular and cellular phenotypes to organismal traits and diseases. 
We focus on evaluations conducted using the types of variants that are typically observed in natural human genetic variation, such as single nucleotide variants (SNVs), as opposed to evaluations that specifically replace or remove whole regulatory elements to probe the models’ learned regulatory grammar. We also focus on non-coding variants in particular, since the potential regulatory impact of coding variants can be difficult to disentangle from downstream impact related to the affected protein. Our comprehensive review of how genomic deep learning models can be leveraged for non-coding variant effect prediction is intended as a resource for both the variant interpretation and the machine learning for genomics communities. To that end, we describe practical considerations for making non-coding variant effect predictions, highlight key insights from current model performance evaluations on non-coding variant effect prediction tasks, and discuss emerging directions for model improvement.

\section*{Making variant effect predictions with sequence-based models}\label{sec:makingVEPs}

\begin{figure}[b]
\centering
\includegraphics[width=\linewidth]{Figure_1_new.png}
\caption{\textbf{Overview of genomic deep learning modeling paradigms and variant effect prediction.} a) Supervised sequence-to-activity models are trained to predict functional genomics readouts from one-hot encoded DNA sequence. b) Self-supervised genomic language models are typically trained to predict masked portions of the input DNA sequence from the surrounding unmasked sequence context. c) Functional-output-based variant effect predictions compare biologically-meaningful output predictions from reference and alternate allele sequences, with appropriate aggregation across tasks. d) Embedding-based variant effect predictions utilize intermediate representations from a supervised or self-supervised model to compare reference and alternate allele sequences. e) In the purely self-supervised setting, likelihood-based variant effect predictions can be made by comparing predicted likelihoods of reference and alternate alleles at the variant location. Self-supervised models can also be fine-tuned on supervised tasks (\ref{sec:box3}) and used to make functional-output-based variant effect predictions as in (c).}
\label{fig:overview}
\end{figure}

Sequence-based deep learning approaches, adapted from computer vision and natural language processing, have shown success in modeling genomic sequences and functional genomics data \cite{Angermuller2016, Zou2018, Eraslan2019, Wang2024review, sasse2024unlocking, sokolova2024deep, capitanchik2024computational, Consens2025, Benegas2025LMreview}. Two modeling paradigms, supervised and self-supervised, are commonly used (Fig. \ref{fig:overview}a,b).
Both types of models can make predictions for any DNA sequence of the appropriate input size, enabling prediction on arbitrary variant-containing sequences.
Supervised sequence-to-activity models (also called sequence-to-function models) 
learn to predict functional activity levels, or the measurements obtained from genome-wide experimental assays of molecular phenotypes (such as chromatin accessibility or histone modifications), from DNA sequence alone (Fig. \ref{fig:overview}a, \ref{sec:box1}). 
In contrast, self-supervised genomic language models (Fig. \ref{fig:overview}b, \ref{sec:box2}) \cite{Consens2025, Benegas2025LMreview} use tasks adapted from natural language processing, such as learning to predict a masked portion of the input DNA sequence from just the surrounding sequence context, to learn representations of DNA sequences.
Both self-supervised and supervised models can subsequently be trained (see \ref{sec:box3} for probing and fine-tuning approaches) on labeled data from supervised tasks of interest, such as additional genome-wide experimental assays or variant effect measurements.

To make a variant effect prediction using a supervised model, two separate functional activity predictions---one for the sequence containing the reference allele and one for the alternate allele---are generated and compared (Fig. \ref{fig:overview}c).
Different distance metrics can be used for this comparison, including absolute (e.g. difference of counts) or relative (e.g. log fold-change) differences between the reference and alternate predictions \cite{Zhou2018, Avsec2021a}.
Many supervised models use multi-task learning to make simultaneous predictions for hundreds or thousands of cell types and experimental assays, resulting in variant effect predictions for each of these tasks. When multiple tasks are relevant, their effect predictions can be aggregated into a single score per variant using several approaches (Fig. \ref{fig:overview}c), such as averaging predictions over many cell types or assays \cite{Kelley2016, Zhou2019}, taking the maximum prediction \cite{Dey2020}, utilizing a lower-dimensional representation of the predictions \cite{Chen2022}, or training an additional supervised model (e.g. the probing approach in \ref{sec:box3}) to learn an appropriate weighting over the predictions for a downstream task \cite{Zhou2015, Kelley2018, Avsec2021b}.
Another choice, particularly for long-context models, is where along a genomic sequence to predict the effect of a variant. For example, a variant might have a proximal effect on chromatin state while also modulating expression of a distal gene. Effect predictions can be computed at the genomic location of the variant itself, at a gene start site or other site of interest (assuming the distance to the variant is less than the model input size), or aggregated over a larger genomic region \cite{Avsec2021b, Linder2025}. The input sequence can also be centered at either the variant or another site such as a gene start site.
Finally, several recent works have reported performance improvements on variant effect prediction tasks from ensembling \cite{bajwa2024characterizing, Linder2025, Lal2024, Avsec2025alphagenome} and distilling \cite{Zhou2024distill, Avsec2025alphagenome} multiple of the same model trained on different random seeds.

For purely self-supervised models, which are not trained on labeled functional data, unsupervised variant effect predictions can be made in several ways (Figure \ref{fig:overview}d,e, \ref{sec:box2}).
One approach is to compare sequence embeddings, which are the model's internal representations of input sequences, for the reference and alternate allele-containing sequences (Fig. \ref{fig:overview}d). Different distance metrics in embedding space have been proposed \cite{DallaTorre2024, Chen2024, marin2024bend}.
Another common method for making variant effect predictions from genomic language models involves masking the variant location within the input sequence and obtaining the predicted probability of each allele being found at that position based on the surrounding sequence. The predicted probabilities for reference and alternate alleles (or allele-containing tokens; see \ref{sec:box2}) can be compared using simple distance metrics such as the log-likelihood ratio (Fig. \ref{fig:overview}e) \cite{Benegas2025gpnmsa}.
One recently proposed metric also quantifies how much a variant influences a model's predicted probabilities for surrounding nucleotides \cite{TomazdaSilva2025}, since functional variants are likely to impact the functional interpretation of the surrounding sequence.
Finally, for self-supervised models that have undergone subsequent additional training on labeled data (probing or fine-tuning; \ref{sec:box3}), variant effect predictions can be made as they would be for a supervised model  (Fig. \ref{fig:overview}c).

We note that key decisions in constructing variant effect predictions, such as distance metrics and the location and aggregation of predicted effects, can yield different results and have not yet been systematically compared.
The best methods for computing variant effects may be specific to particular models or tasks, but differences in these choices can make direct comparison of reported performance values difficult.

\section*{Validation data and task construction}

\subsection*{Sources of validation data} 
To validate variant effect predictions, two broad categories of variant effect data are typically used as ground truth: observational data on naturally occurring human genetic variation, and experimental data on introduced variation in laboratory contexts (Fig. \ref{fig:data}).
 
Association studies of observed genetic variants, such as GWAS and quantitative trait locus (QTL) studies, correlate variant dosages with organism-level phenotypes or quantitative molecular traits such as gene expression levels, relying on data from large human cohorts \cite{uffelmann2021genome, Aguet2023}. These studies produce summary statistics quantifying the marginal effect of each variant on the phenotype, although differences in methodology for computing these effect sizes can affect their biological interpretation \cite{Aguet2023}. 
A limitation is that these studies identify both causal and non-causal variant associations due to correlations between nearby variants, or LD. Statistical fine-mapping methods \cite{schaid2018genome} can reduce, but not eliminate, the number of non-causal associations. Furthermore, association studies have power to detect effects of only relatively common variants, and they are biased by the genetic ancestry of studied cohorts. These limitations contribute to label errors in evaluation sets constructed from association study data, lowering the maximum achievable variant effect prediction performance.

Related types of observational variant effect data include rare variants associated with extreme or outlier phenotypes \cite{li2017river, ferraro2020watershed}, which tend to have large effect sizes,
and allelic imbalance data quantifying allele-specific activity at heterozygous variant sites. Allelic imbalance data can be obtained from even a single individual, but are limited to variants that are heterozygous in the sample and fall within sequenced regions of the assay.

Another source of observational data is variant databases such as ClinVar \cite{Landrum2017} and the Human Gene Mutation Database (HGMD) \cite{Stenson2014} that contain categorical variant  
labels such as pathogenic or benign. Because these labels primarily reflect variant impact in the context of rare disease, e.g. based on clinical classification guidelines \cite{Richards2015acmg}, the pathogenic or functionally significant variants in these databases are predominantly rare and have much larger effects than GWAS-identified variants. They also tend to be coding variants, though large-effect non-coding variants are also present. Unlike association study data, these databases do not suffer from linkage-based non-causal associations, but can contain label errors due to other challenges in clinical variant interpretation. They are also biased towards variants occurring in well-studied diseases and genes. 

Experimental assays, such as massively parallel reporter assays (MPRAs) and CRISPR-based perturbation experiments, test functional activity levels of synthetic DNA sequences or make perturbations to native DNA, enabling the collection of functional data from sequences not present in observed populations \cite{Rong2024}. Both assay types provide quantitative estimates of effect. MPRAs assay the effects of short sequences on a given regulatory phenotype, such as transcription, splicing, polyadenylation, or ribosome load \cite{Tewhey2016, adamson2018vexseq, WeingartenGabbay2019, Kircher2019, Bogard2019, Bergman2022, Siraj2024}, and measure activity from many sequences in a single high-throughput experiment. MPRAs have been used to compare sequences containing human reference and alternate alleles, perform saturation mutagenesis of regulatory elements, or test random sequences. 
However, MPRAs do not reflect functional activity within the native chromatin context.
CRISPR-based assays --- such as CRISPR interference (CRISPRi), in which regulatory activity from targeted regions is suppressed, and CRISPR prime editing, which enables precise sequence edits with up to single base resolution --- can measure the effects of perturbations in a native chromatin context within the endogenous genome \cite{komor2017crispr}, and continue to improve in efficacy, resolution, and throughput.

\subsection*{Evaluation task construction}
Different types of tasks can be constructed from the variant effect datasets discussed above, providing complementary information about model performance. 

Commonly used tasks include: (1) functional variant identification, a binary classification task that evaluates whether models can distinguish a positive set of functional variants (e.g. significantly associated with a change in activity) from a negative set of non-functional variants;  
(2) effect direction prediction, another binary classification task that evaluates whether models can predict whether a variant increases or decreases a particular phenotype;  
and (3) effect size prediction, a regression task that evaluates whether models can quantitatively predict the effect size of a variant, either signed (testing prediction of both magnitude and direction) or unsigned (magnitude only).
When genome-wide summary statistics from GWAS or QTL studies are used to evaluate predicted effect sizes, approaches such as stratified LD score regression \cite{finucane2015ldsc} and signed LD profile (SLDP) regression \cite{Reshef2018}, which explicitly account for LD, can also be used \cite{Dey2020, Avsec2021b, murphy2024predicting}.
Accuracy on each type of task may provide different information about a model's utility for biologically-relevant tasks. For example, while good performance on identification-type tasks is promising for downstream applications related to causal variant identification, such as functionally-informed fine-mapping or clinical variant interpretation, predicting the direction of effect of a disease-associated variant can provide mechanistic hypotheses, such as whether up- or down-regulation of a gene mediates disease.

Several aspects of evaluation task construction can bias our understanding of model accuracy. For functional variant identification tasks, different choices of negative set can yield vastly different results. For example, if the non-functional variants are chosen from non-functional genomic regions, models can excel simply by distinguishing functional from non-functional regions, an easier task than distinguishing variants impacting function from non-functional variants within the same region or regulatory element. Similarly, if a positive set consists of annotated pathogenic variants, which typically have low population allele frequencies, negative sets can be matched on allele frequency to ensure that simply distinguishing rare from common variants is insufficient to obtain high performance.  
Furthermore, for all evaluation tasks, any choices that affect the distribution of effect sizes for variants included in the evaluation dataset (for example, using variants from association studies with different power, or using different significance thresholds to select variants) will impact the overall difficulty of the task and affect reported performance, making performance metrics hard to compare across studies even for similar-sounding evaluations. 
Careful consideration of the details of evaluation task construction and any matching that was performed is therefore essential to correctly interpret and compare results between studies.

\section*{Evaluations of variant effect predictions}\label{sec:evalVEP}

\begin{figure}
\centering
\includegraphics[width=\linewidth]{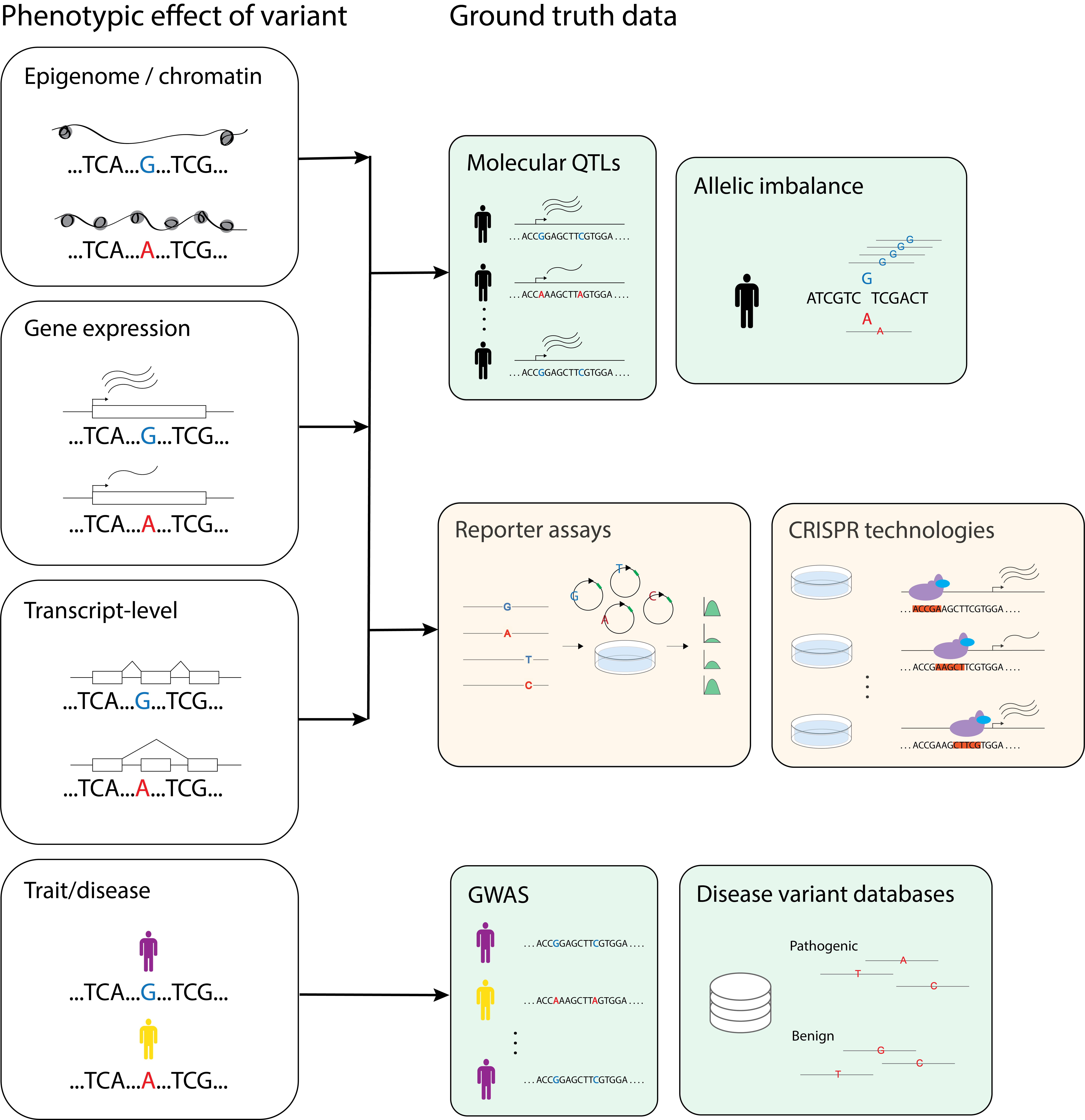}
\caption{\textbf{Variant effects at multiple scales with diverse ground truth data.} Variant effects can be measured on a range of phenotypes (left), including epigenetic or chromatin states, gene expression levels (transcription), transcript-level features, and disease or complex traits. Observational (right, green) or experimental (right, yellow) data can be used as ground truth to evaluate variant effect prediction accuracy for each of these phenotypes.
}
\label{fig:data}
\end{figure}

Here we review evaluations that have been performed to date using genomic deep learning models for non-coding variant effect prediction. The effect of a variant can be observed across multiple regulatory scales, including effect on epigenetic or chromatin state, gene expression, transcript-level regulation, and organism-level trait or disease (Fig. \ref{fig:data}).
Successful variant effect prediction at a given scale may require specific modeling choices to capture distinct biology; for example, chromatin states tend to be modulated by proximal sequence \cite{degner2012dnase}, while gene expression can be regulated by distal enhancers \cite{Gasperini2020}.
In addition, non-coding variant effects are often specific to particular cell types or states, necessitating accurate modeling of cell-type-specific effects.
We discuss and synthesize existing evaluations of variant effect prediction at different regulatory scales, including evaluations performed in the original modeling papers as well as those performed in separate benchmarking studies (see also Supplementary Table 1).

\subsection*{Predicting variant effects on epigenetic or chromatin state}
Local epigenetic or chromatin state includes DNA methylation, histone modifications, transcription factor (TF) binding, and accessibility.
Many supervised sequence-to-activity models are trained directly on such phenotypes; however, predicting a variant's effect on these phenotypes requires a causal understanding of relevant sequence features, such as TF binding motif disruption.

Multiple studies have demonstrated the utility of supervised sequence-to-activity models, both multi-task and single-task, for predicting variant effects on local chromatin state—--including functional variant identification, effect direction, and effect size prediction tasks---using chromatin accessibility, methylation, histone modification, and TF binding variant effect datasets \cite{Zhou2015, Alipanahi2015, Zeng2017cpgenie,
Wagih2018, Hoffman2019deepfigv, Han2024, PampariChrombpnet2024, Avsec2025alphagenome}. 
In addition, a few structural variants with experimentally-validated effects on genome organization are correctly predicted by supervised models trained on 3D chromatin structure \cite{Fudenberg2020, Schwessinger2020, Trieu2020deepmilo, Zhou2022}, but larger datasets of such variants are needed for systematic evaluation. A consistent finding is that predicted variant effects are more accurate when the magnitude of the predicted difference between reference and alternate alleles is larger, as demonstrated by DeepSEA for chromatin accessibility and histone modification variants \cite{Zhou2015}, CpGenie for methylation QTLs (meQTLs) \cite{Zeng2017cpgenie}, and DeepFIGV for chromatin accessibility QTLs \cite{Hoffman2019deepfigv}. Another finding is that small models tend to perform well when predicting variant effects on local chromatin state. Single-task ChromBPNet models (utilizing $\sim$2kb sequences) perform similarly to the large multi-task model Enformer (utilizing $\sim$200kb sequences) on accessibility QTL identification and effect size predictions \cite{PampariChrombpnet2024}, suggesting that distal sequence context does not improve prediction of variant effects on accessibility. The ChromBPNet authors were careful to select negative set variants that fall in functional regions but do not impact function \cite{PampariChrombpnet2024}, a more challenging choice of negative set than variants in non-functional regions.
The large multi-task model AlphaGenome (utilizing $\sim$1Mb sequences) reports improved performance over previous models on accessibility QTLs, but notes only minor benefit from the long sequence context and no benefit from multi-task training on additional data types beyond accessibility \cite{Avsec2025alphagenome}.

The utility of self-supervised models for predicting variant effects on local chromatin state remains unclear.
One comparison found that unsupervised, probed, and fine-tuned versions of several self-supervised models
performed worse than ChromBPNet on an accessibility QTL identification task, despite fine-tuned models matching ChromBPNet's performance on accessibility prediction for reference sequences \cite{patel2024darteval}.
In addition, the Nucleotide Transformer authors reported that their self-supervised models did not outperform an evolutionary conservation-based score \cite{Cooper2005} on an meQTL identification task, even when fine-tuned \cite{DallaTorre2024}.

\subsection*{Predicting variant effects on gene expression}
Many genomic deep learning models have been evaluated on their ability to predict variant effects on gene expression and generally perform better than non-deep-learning baselines \cite{Zhou2015, quang2016danq, Zeng2017cpgenie, Zhou2018, Kelley2018, wang2018define, Hoffman2019deepfigv, Fudenberg2020, Schwessinger2020, minnoye2020cross, Avsec2021b, Chen2022, Toneyan2022, Karollus2023distal, celaj2023bigrna, Hlzlwimmer2025abexp, Lal2024, murphy2024predicting, hingerl2025scooby, Linder2025, Jaganathan2025promoterai, Avsec2025alphagenome}. Several trends have emerged from these evaluations.
Supervised sequence-to-activity models tend to perform well for identifying gene-proximal expression-altering variants and predicting their directions of effect.
However, models show room for improvement in predicting distal variant effects, quantitative effect sizes, and personalized gene expression, as discussed below. 
While most evaluations focus on SNVs, the AlphaGenome authors report only slightly lower performance for indels on several eQTL prediction tasks \cite{Avsec2025alphagenome}.
Evaluations of self-supervised models have been more limited, but they often fail to match supervised model performance. For example, one independent evaluation reported that unsupervised predictions from many self-supervised models underperformed DeepSEA predictions on an eQTL identification task \cite{marin2024bend}.

A major advance in sequence-to-expression modeling has been incorporating longer sequence contexts, and comparisons of the Basenji2 \cite{Kelley2020}, Enformer \cite{Avsec2021b}, Borzoi \cite{Linder2025}, and AlphaGenome \cite{Avsec2025alphagenome} models---which have roughly 50kb, 200kb, 500kb, and 1Mb sequence contexts, respectively---show that longer context models are often better at identifying eQTLs and predicting their directions of effect \cite{Avsec2021b, Linder2025, Avsec2025alphagenome}. However, these models struggle to predict distal variant effects, even for variants within their receptive fields. In particular, eQTL evaluations stratified by distance to the transcription start site (TSS) show that Basenji2, Enformer, Borzoi, and AlphaGenome have much stronger performance for gene-proximal variants (within a few kilobases of the TSS) than distal variants \cite{Avsec2021b, Linder2025, Avsec2025alphagenome}. Strong proximal variant performance is also demonstrated in recent analyses using rare variants associated with gene expression outliers, which found that tested models---including Enformer, Borzoi, and PromoterAI---can accurately predict the direction of effect for variants within a few hundred basepairs of the TSS \cite{Hlzlwimmer2025abexp, Jaganathan2025promoterai}. 

Distal variants tend to have smaller effects than proximal variants, and accurate prediction of gene expression effects  requires models to learn the effect of distal regulatory regions on their target genes. Several studies using eQTL or expression MPRA evaluation data have reported more accurate direction prediction for variants with larger predicted magnitudes of effect \cite{Zhou2015, Zhou2018, hingerl2025scooby, schwessinger2023seq2cells, Avsec2025alphagenome}. Another study reported high uncertainty in eQTL direction predictions from independently trained Basenji2 models, particularly for variants with smaller predicted effects \cite{bajwa2024characterizing}. 
These results suggest that better performance on variants with large effects may contribute to the better performance observed for gene-proximal variants, though different biology for gene-proximal and gene-distal variants is likely also an important factor.
Several CRISPR-based evaluations show that models struggle to link the effects of distal regulatory regions to their target genes, likely contributing to poor distal eQTL prediction. Basenji2, Enformer, Borzoi, and AlphaGenome identify CRISPRi-validated enhancer-gene pairs \cite{Gasperini2019, Fulco2019} more accurately for proximal enhancers than distal enhancers \cite{Avsec2021b, Linder2025, Avsec2025alphagenome}, and predicted enhancer effects underestimate experimentally-measured effects for distal enhancers \cite{Karollus2023distal}. Another study used CRISPR prime editing to measure effects on expression of tiled 5-bp edits of a specific promoter and distal enhancer \cite{Martyn2025} and found that for edits in the distal enhancer, the correlation with experimentally-measured expression effects was very low for Enformer effect predictions at the TSS. However, utilizing accessibility effect predictions at the enhancer---from either Enformer or ChromBPNet---scaled by the experimentally-measured effect of the enhancer on overall expression achieved a moderate correlation for gene expression effect prediction, suggesting that models predict local effects well, but fail to learn the scaling required to correctly link distal elements to target genes.
In contrast, for the CRISPR-induced promoter edits, all tested models---Enformer, ChromBPNet, and ProCapNet \cite{Cochran2024}---showed moderate correlation between experimental measurements and predictions \cite{Martyn2025}.

Prediction of quantitative variant effect sizes, for which both sign and magnitude are important, can also be improved. Several studies report moderate performance on this task \cite{Kelley2018, Hoffman2019deepfigv, minnoye2020cross, Avsec2021b, Karollus2023distal, Chen2022, Toneyan2022, Linder2025, Lal2024, hingerl2025scooby, murphy2024predicting, Jaganathan2025promoterai, Avsec2025alphagenome}, though AlphaGenome notes that unsigned correlations (magnitude only) between model predicted effects and fine-mapped eQTL effects are much lower than signed correlations \cite{Avsec2025alphagenome}, suggesting that accurate direction-of-effect prediction for variants with large magnitudes of effect may drive reported signed correlations in many cases.
Effect size correlations reported by Borzoi varied substantially between tissues \cite{Linder2025}. Enformer's reported correlations with MPRA-measured variant effect sizes \cite{Kircher2019} were highly variable for different promoters and enhancers, and all correlations were substantially lower than the theoretical limit given by experimental replicate correlations \cite{Avsec2021b}.
Several studies have noted that MPRA-measured effect sizes correlate better with predicted variant effects on local accessibility than on gene expression \cite{Avsec2021b, Karollus2023distal, PampariChrombpnet2024}, which may suggest that either the modeling choices or the MPRA biology itself is more suited to accessibility prediction.
Finally, the CLIPNET authors report moderate effect size correlations using their local-context model to predict the effect sizes of transcription initiation QTLs \cite{he2024dissection}, consistent with locally-regulated effects being easier to model than distal effects.

A final area for improvement is predicting inter-individual variation in gene expression from personal genome sequences, a challenging task that requires models to identify which proximal and distal variants within a personal genome modulate gene expression and to correctly predict the direction and magnitude of their effects. Two studies tested the ability of supervised models to predict gene expression from personal genomes and reported poor performance when predicting inter-individual variation in gene expression for most genes \cite{Sasse2023, Huang2023}. For this type of evaluation, predictions are made at the gene TSS, which requires predicting distal effects for many of the personal genome variants, and are combined across the two phased haplotypes for each individual. Follow-up work \cite{Drusinsky2024,Rastogi2024, spiro2025scalable} found that fine-tuning models on paired personal genome and transcriptome data leads to improvement on this task, with performance comparable to linear variant-based models \cite{Gamazon2015, gusev2016integrative}, but only for genes seen during fine-tuning. The poor performance on unseen genes suggests that this strategy does not yet generalize to unseen (rare or \textit{de novo}) variants. However, independent work suggests that training on personal genome sequences can improve prediction of variant effects from local-context transcription initiation models \cite{he2024training} and that training on sequences containing rare variants associated with gene expression outliers can improve prediction of TSS-proximal variant effects \cite{Jaganathan2025promoterai}.

\subsection*{Predicting variant effects on transcript-level regulation}
Disease-associated variants can act by altering transcript-level regulatory mechanisms, such as those involved in splicing, polyadenylation, and RNA stability. Many genomic deep learning models, often designed specifically for these phenotypes, have been evaluated on their ability to predict transcript-level variant effects.

\textbf{Splicing}. Supervised models trained to predict RNA splicing patterns from DNA sequence have been tested on their ability to predict variant effects on splicing \cite{Jaganathan2019, Cheng2019, Cheng2021, Zeng2022, celaj2023bigrna, Chen2024, Xu2024deltasplice, You2024splicetransformer, Linder2025, Avsec2025alphagenome}. These studies collectively reveal that sequence-based models are highly predictive of splicing effects for splice-site proximal variants, but struggle with distal variants and tissue-specific variants \cite{Jaganathan2019, Zeng2022, wagner2023absplice, Linder2025}.
For example, Pangolin reported a substantial drop in performance for variants $>9$ bases from a splice site, evaluated using experimental splicing assays \cite{Zeng2022}.
Two models with different training strategies -- Pangolin and Borzoi -- have different relative strengths on proximal and distal variants when identifying distance-stratified splicing QTLs (sQTLs) \cite{Linder2025}, suggesting that different modeling strategies in these regimes may be optimal.
Another evaluation found that tested models -- MMSplice and SpliceAI -- have limited ability to predict the tissue-specific effects of rare variants associated with aberrant splicing outlier events \cite{wagner2023absplice}.
AlphaGenome reported improved performance on these aberrant splicing outlier variants, including both SNVs and indels, by explicitly modeling splice sites, splice site usage, and splice junctions together with other data modalities \cite{Avsec2025alphagenome}.

Self-supervised models have been evaluated for splicing variant effect prediction in a few cases \cite{Ji2021, brixi2025evo2}.
 SpliceBERT, a self-supervised model trained on multi-species RNA sequences, outperformed DNABERT \cite{Ji2021}, a self-supervised model trained on the entire human genome, for identifying splice-disrupting variants \cite{Chen2024}.
 More studies are needed to establish whether self-supervised models trained on entire genomes---which attempt to learn salient aspects of DNA, RNA, and protein regulation within one unified model---outperform specialized models tailored to specific aspects of genome biology.

\textbf{Other transcript-level regulation.}
Several supervised models have been evaluated on their ability to predict effects of non-coding variants in 3' and 5' untranslated regions (UTRs). These tend to be specialized models trained to predict biological processes regulated by UTR sequences, such as mean ribosome load (a proxy for translation rate) for 5' UTRs and mRNA half-life and polyadenylation site (PAS) usage for 3' UTRs. 

For 5' UTR variants, Optimus 5-Prime \cite{sample2019optimus}---trained on a random sequence MPRA measuring mean ribosome load---accurately predicted MPRA-measured effects of 5' UTR SNVs from ClinVar. The Framepool \cite{Karollus2021framepool} authors extended these modeling capabilities to predict variant effects in 5’ UTRs of longer lengths. Notably, these studies used random, synthetic sequences as training data, allowing the models to learn from greater sequence diversity than that of endogenous sequences. This strategy has the potential to improve variant effect prediction \cite{deBoer2023} but has not yet been systematically assessed.

For 3' UTR variants, models predicting mRNA half-life and PAS usage can accurately predict variant effects on these same phenotypes. The Saluki mRNA half-life model predicted RNA stability measurements from saturation mutagenesis MPRA experiments on par with the theoretical limit given by experimental replicate correlations \cite{Agarwal2022}. Several supervised models that predict PAS usage from sequence are highly accurate for predicting the effect of PAS-proximal variants on polyadenylation \cite{Bogard2019, Arefeen2019, VainbergSlutskin2019, Li2021, Linder2022}. Polyadenylation variant effect prediction accuracy decreases markedly with distance outside of the core hexamer polyadenylation motif \cite{Linder2022, Linder2025, Avsec2025alphagenome}, though Borzoi \cite{Linder2025} and AlphaGenome \cite{Avsec2025alphagenome} report substantial improvements in accuracy for PAS-distal variants compared to prior models.

\subsection*{Predicting variant effects on trait and disease}

Many genomic deep learning models can identify variants involved in organism-level traits and diseases, despite never seeing such annotations during training.

\textbf{Rare disease variants}. Variants involved in rare disease typically have large effects on downstream phenotypes, and some models achieve moderate performance---comparable to the non-deep-learning predictor CADD \cite{kircher2014general}---for identifying non-coding variants implicated in rare disease \cite{Zhou2015, Chen2022, Hoffman2019deepfigv, Cheng2019, Zeng2022, Leung2018, celaj2023bigrna, Ji2021, Yang2022, DallaTorre2024, Benegas2025gpnmsa, brixi2025evo2, Jaganathan2025promoterai, Avsec2025alphagenome}. Some supervised models have particularly strong performance on certain classes of non-coding variants that tend to have large phenotypic impact, such as disease-relevant variants from ClinVar located close to splice sites \cite{Cheng2019, Zeng2022}, PASs \cite{Leung2018}, or within UTRs \cite{celaj2023bigrna}. Self-supervised models are competitive with, and sometimes outperform, supervised models for identifying disease-relevant variants from curated variant databases \cite{Ji2021, Yang2022, marin2024bend, benegas2025traitgym},  particularly when they learn information related to evolutionary conservation by training on sequences from many species or on multiple sequence alignments \cite{benegas2025traitgym}. An independent study found that, on a non-coding ClinVar variant identification task, the multi-species Nucleotide Transformer model outperformed DeepSEA and self-supervised models trained only on human data \cite{marin2024bend}. The multiple sequence alignment-based self-supervised model GPN-MSA \cite{Benegas2025gpnmsa} also outperformed supervised models including Borzoi for identifying non-coding variants involved in Mendelian diseases \cite{benegas2025traitgym}. 
We note that many reported evaluations on disease-causing variants, even ones utilizing the same underlying database, can be difficult to directly compare due to varying choices in which variants to include and how to construct negative variant sets. For example, for matched negative variants, Nucleotide Transformer selected common variants within 100kb of HGMD pathogenic variants \cite{DallaTorre2024}, while LOGO selected common variants within just 1kb of the pathogenic variants \cite{Yang2022}. Different negative sets have been shown to substantially impact model performance \cite{benegas2025traitgym, Avsec2025alphagenome}. Finally, rare disease variant databases contain primarily coding variants, and some reported evaluations include both coding and non-coding variants, making it difficult to disentangle performance on these respective variant classes.

\textbf{Complex trait variants}. Understanding common variants with small phenotypic effects is important for complex traits and diseases, including GWAS interpretation. Both supervised and self-supervised models demonstrate moderate ability to identify trait-associated GWAS variants \cite{Zhou2015, Kelley2016, Kelley2018, Yang2022, PampariChrombpnet2024, Sokolova2023, Linder2025, Avsec2025alphagenome}, but recent work suggests that supervised models are generally superior in this setting \cite{benegas2025traitgym}.
Evaluations vary in the choice of GWAS dataset, whether fine-mapping approaches were applied, and how negative variants were chosen, which all impact reported performance. On a GWAS variant identification dataset with carefully
matched negative variants \cite{benegas2025traitgym}, AlphaGenome, Borzoi, and Enformer had comparable performance to CADD \cite{benegas2025traitgym, Avsec2025alphagenome}. While DeepSEA and Basenji outperformed CADD and other non-deep-learning baselines on a different GWAS variant identification dataset \cite{Zhou2015, Kelley2018}, a complementary study tested the conditional informativeness of DeepSEA and Basenji variant effect predictions for stratified LD score regression \cite{finucane2015ldsc} when existing coding, conservation, regulatory element, and LD-related annotations were considered, finding that model predictions provided little information beyond these existing annotations, despite being highly enriched for trait heritability \cite{Dey2020}. 
Together, these results suggest that current models have learned signals important for identifying complex trait-relevant variants, but further work is needed to establish whether they can augment the rich signals already present in existing evolutionary and functional datasets.

Finally, expanding the cell types and states profiled in functional assays, and training models on these datasets, will be important for trait-relevant variant effect prediction, since variant effects on organism-level traits and diseases are often mediated by effects on molecular phenotypes that occur only in specific cellular contexts, making large and diverse datasets crucial.

\section*{Discussion}\label{discussion}

In this Review, we discuss leveraging genomic deep learning models for non-coding variant effect prediction.
Supervised sequence-to-activity models can identify and provide mechanistic interpretations of non-coding variants impacting a range of molecular phenotypes. Current models typically perform best for variants with high effect sizes and for local effects close to the site of a variant. 
There is emerging consensus that we can predict local regulatory activity fairly well, while predicting the impact of regulatory activity on distally regulated phenotypes such as gene expression remains a harder problem, even for the most recent and widest context models.
Biologically, enhancer-promoter and enhancer-enhancer interactions define a complex $cis$-regulatory logic that we do not fully understand \cite{Thomas2025}. Deep learning approaches are data-hungry, and it is possible that functional activity measurements from the available set of endogenous human genes may be insufficient to learn the higher order $cis$-regulatory rules empirically \cite{deBoer2023}. While advances in model architectures, resolution, and compute have led to current state-of-the-art performance, additional gains may require updated modeling paradigms for utilizing additional biological knowledge.

Promising approaches to better model higher order sequence grammar include models that take into account additional experimental data beyond sequence \cite{Lin2024epinformer} or utilize biologically motivated inductive biases, such as interpretable-by-design models \cite{balci2023intrinsically, Dudnyk2024}.
Single cell data has also become increasingly abundant. Models trained on single cell data \cite{Yuan2022, schwessinger2023seq2cells, hingerl2025scooby} and approaches to predict context-specific effects without explicit training data \cite{nair2019integrating, hingerl2025scooby, murphy2024predicting} may be effective for predicting context-specific variant effects in cell types for which we have little or no experimental data.

Self-supervised genomic language models have been less comprehensively evaluated on non-coding variant effect prediction than supervised models, though they achieve similar performance on some tasks when fine-tuned \cite{DallaTorre2024, patel2024darteval, benegas2025traitgym}. Their practical utility relative to supervised models remains unclear from recent benchmarking studies \cite{Tang2025, marin2024bend, kao2024lrbenchmark, patel2024darteval, Cheng2025longbench, benegas2025traitgym}.
Self-supervised training has been transformative in modeling natural language, a domain which has vast amounts of unlabeled data and smaller amounts of labeled data. In contrast, the human genome is not an unlimited source of sequence data, and genome-wide functional activity labels are available for many important molecular phenotypes. Therefore, it is unclear whether self-supervised training will provide the same benefits it does for modeling natural language. Due to sequence conservation of important regulatory elements and regulatory grammar, leveraging genomes from multiple species \cite{DallaTorre2024, Benegas2025gpnmsa} can increase the amount of unlabeled training data available, potentially leading to better performance on human tasks. However, sequence similarity between species also limits the diversity gained.
The relative success of multi-species self-supervised models on predicting the large effects of rare and disease variants is likely due to them effectively learning about sequence conservation and fitness. This hypothesis suggests that supervised model predictions, informed by the mechanisms of gene regulation, may be best utilized alongside orthogonal measures of conservation or gene constraint for prioritizing trait and disease variants \cite{marderstein2025mapping}.

All models are fundamentally limited by the amount of sequence diversity available for training, which constrains their ability to understand the effects of small sequence variations.
The model design choices that enable good performance on reference sequence training data may be suboptimal for variant effect prediction \cite{Kathail2024}, highlighting the importance of considering variant-based training and evaluation.
In particular, future work should systematically explore training models on data from a variety of MPRA designs, such as random sequences \cite{deBoer2023} or mutagenized endogenous sequences, to evaluate whether this additional sequence diversity
can improve variant effect prediction in the context of naturally occurring, trait-relevant genetic variation. 
As the scale and throughput of MPRA and CRISPR-based experiments continues to grow, we anticipate that these datasets will become a rich source of training and evaluation data for genomic deep learning models.

In addition to well-maintained and accessible code for running sequence-based models, which is important for their downstream application in clinical variant interpretation settings, the adoption of variant effect benchmarks and competitions \cite{Kryshtafovych2023, cagi2024} could better standardize data formats and model comparisons. These formal settings would enable systematic comparison of modeling and variant scoring decisions, a clear current gap in the field, on appropriately challenging and biologically realistic benchmark tasks.

Finally, while existing model evaluations have focused on single-nucleotide variants, insertions, deletions, and structural variants are important contributors to phenotypic variation. Predicting their effects requires comparing sequences of different lengths, and robust evaluations and best practices are still being established \cite{Avsec2025alphagenome, Korsakova2025}.
Sequence-based genomic deep learning models remain a promising approach to non-coding variant effect prediction, and future modeling improvements and systematic benchmarking analyses should continue to improve their utility.

\section*{Acknowledgements}
We thank members of the Ioannidis lab for helpful comments, particularly eyes robson, Ruchir Rastogi, and Aniketh Reddy. This work was partially supported by the UC Noyce Initiative for Computational Transformation. N.M.I. is a Chan Zuckerberg Biohub San Francisco Investigator.

\section*{Author contributions}
All authors contributed to all aspects of the article.

\section*{Competing interests}
The authors declare no competing interests.

\clearpage

\renewcommand{\thesection}{Box \arabic{section}} 
\setcounter{section}{0}

\noindent\makebox[\linewidth]{\rule{0.625\paperwidth}{0.4pt}}
\section{Supervised sequence-to-activity models}\label{sec:box1}

\begin{figure}[b]
\centering
\includegraphics[width=\linewidth]{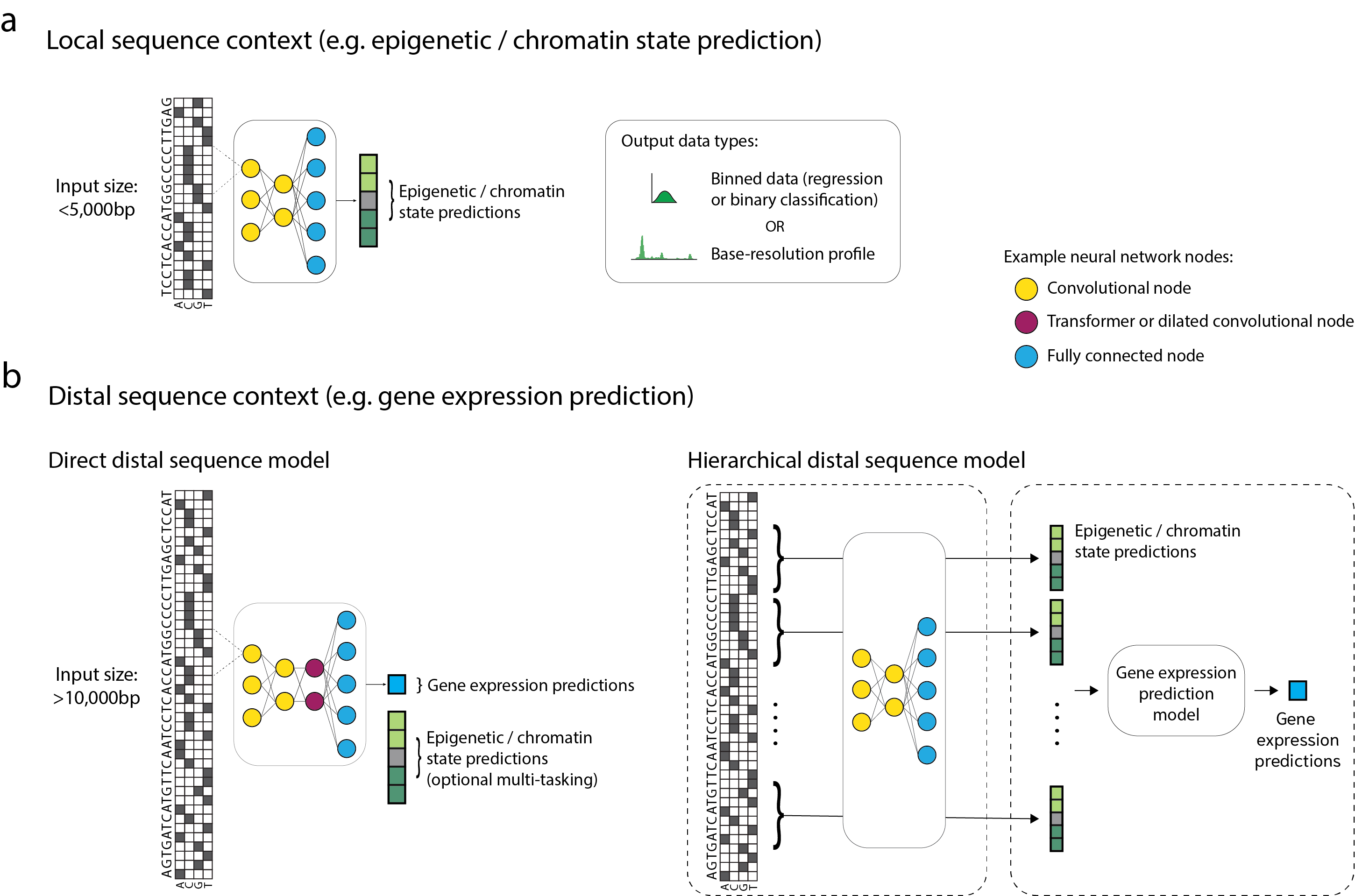}
\caption*{\textbf{Box 1 Figure: Sequence-to-activity model design choices for local and long-context prediction.}}
\label{fig:model_types}
\end{figure}

Supervised sequence-to-activity models are neural networks trained to predict functional activity levels, the molecular phenotypes measured in functional assays such as ATAC-seq, ChIP-seq, or RNA-seq, from input DNA sequences.
Design choices include the number, size and type of layers in the network (e.g. fully connected, convolutional, or transformer layers), input and convolutional filter sizes, activation functions, and loss functions \cite{Zou2018, Eraslan2019}. Most supervised models utilize early convolutional layers, since they are a natural choice for modeling DNA sequence motifs \cite{Zou2018}. Design choices are informed by the prediction task(s), whether they are regression or classification, and how much sequence context (the length of input DNA sequence considered by a model) they require. 
While longer sequence contexts make it possible to learn distal regulatory effects, they can also become computationally prohibitive.

\textbf{Local sequence models.} Many models use short input sequences ($<$5,000bp; Box figure a) to make predictions about local molecular phenotypes, such as chromatin accessibility, TF binding, histone modifications, DNA methylation, transcription initiation, polyadenylation, or splicing, using either single-task or multi-task learning \cite{Alipanahi2015, Zhou2015, quang2016danq, Kelley2016, Zeng2017cpgenie, wang2018define, Leung2018, Hoffman2019deepfigv, Cheng2019, Bogard2019, Arefeen2019, VainbergSlutskin2019, sample2019optimus, Maslova2020, minnoye2020cross, Avsec2021a, Cheng2021, Li2021, Karollus2021framepool, Chen2022, Linder2022, PampariChrombpnet2024, Dudnyk2024, he2024dissection, Cochran2024}.
Early examples include DeepSEA \cite{Zhou2015} and Basset \cite{Kelley2016}, which use convolutional, pooling, and fully connected layers to predict functional activity in multiple cell types.
Prediction tasks may be structured as binary classification (e.g. predicting the presence or absence of a called peak) or regression (predicting a continuous-valued level of activity).
Regression models predict either binned read counts for short sequence windows or base resolution profiles (e.g. read depth at every position within the input sequence).
Pooling layers reduce the dimension of the data representations, enabling models to efficiently handle longer input sequences but also reducing the resolution of the output predictions. 
An alternative approach, used by BPNet \cite{Avsec2021a} and others \cite{PampariChrombpnet2024, Linder2022, he2024dissection, Dudnyk2024, Cochran2024, Jaganathan2025promoterai}, involves forgoing or minimizing pooling layers to achieve base resolution prediction.

\textbf{Distal sequence models.} Models that predict distally regulated phenotypes, such as gene expression, aim to capture the effects of both proximal and distal regulatory elements. Many models incorporate wider sequence context by directly expanding the input sequence length (Box figure b, left) \cite{Kelley2018, Jaganathan2019, Kelley2020, Agarwal2020, Avsec2021b, Zeng2022, Agarwal2022, Linder2025, celaj2023bigrna, Xu2024deltasplice, You2024splicetransformer, Linder2025, Jaganathan2025promoterai, Avsec2025alphagenome}. Basenji \cite{Kelley2018, Kelley2020} utilizes dilated convolutions (containing gaps between the convolutional filter values to scan longer inputs without increasing parameter count) to achieve a 32kb sequence context. 
Basenji's successor, Enformer \cite{Avsec2021b}, incorporates a transformer architecture \cite{Vaswani2017} to increase the sequence context to $\sim$200kb. The transformer's self-attention mechanism, originally developed for modeling natural language, enables learning of relevant features from distal regions within an input sequence.
Another successor, Borzoi \cite{Linder2025}, includes a U-net architecture \cite{Ronneberger2015} to enable higher resolution predictions (32bp, versus 128bp for Basenji and Enformer) and an expanded sequence context of $\sim$500kb. PromoterAI \cite{Jaganathan2025promoterai} and AlphaGenome \cite{Avsec2025alphagenome} are recent, more computationally intensive models that have wide contexts (10kb and 1mb, respectively) yet predict at base resolution.
Some distal sequence models predict other phenotypes, including 3D genome folding (Akita \cite{Fudenberg2020}, DeepC \cite{Schwessinger2020}, Orca \cite{Zhou2022}), mRNA splicing (SpliceAI \cite{Jaganathan2019}, Pangolin \cite{Zeng2022}), and transcription initiation (Puffin-D \cite{Dudnyk2024}).
While the largest models have reached context sizes of hundreds of kilobases, they may not capture the full biological context relevant to a given task, such as very distal enhancers. Furthermore, they may not fully utilize distal features within their own input contexts \cite{Karollus2023distal}.

\textbf{Hierarchical distal sequence models.} Rather than directly modeling long input sequences, Expecto \cite{Zhou2018} and ExpectoSC \cite{Sokolova2023} process longer sequences by making epigenetic and chromatin state predictions on shorter sub-sequences, then using these predictions as inputs to secondary models of gene expression (Box figure b, right).
This hierarchical approach is motivated by the biology of gene expression and presents another method for integrating distal information.

Supervised architectures provide a powerful approach to learn salient sequence patterns from functional data. However, they require training data from cell types of interest and cannot be generalized to unseen assays or cell types without additional training or fine-tuning (see also \ref{sec:box3}). 

\vspace{2mm}
\noindent\makebox[\linewidth]{\rule{0.625\paperwidth}{0.4pt}}

\clearpage
\noindent\makebox[\linewidth]{\rule{0.625\paperwidth}{0.4pt}}
\section{Self-supervised genomic language models}\label{sec:box2}
Genomic language modeling is a relatively new paradigm, and many such models trained on human or cross-species genomes have been proposed \cite{Zaheer2020, Ji2021, Yang2022, zhou2024dnabert, DallaTorre2024, Benegas2025gpnmsa, Nguyen2023, fishman2025genaLM, Gndz2023, robson2024guanine, sanabria2023grover, zhang2023dnagpt, schiff2024caduceus, Chen2024, chu2024utrLM, zhu2024CDGPT, McNally2024allspice, shearer2024lol-eve, ellington2024aidodna, brixi2025evo2, Karollus2024}. 
By learning the statistical distributions present in natural genome sequences using self-supervised pre-training tasks derived from the sequences themselves, genomic language models synthesize salient vector representations of sequence features, known as embeddings, during pre-training.
Direct use of the embeddings or predicted nucleotide probabilities (known as unsupervised or zero-shot prediction) may result in good performance on certain sequence-based prediction tasks \cite{Benegas2025gpnmsa, DallaTorre2024}. More commonly, a pre-trained model is taken as a good initialization of model parameters, and the model is further trained on supervised prediction tasks using task-specific datasets (e.g. functional data). This approach may require less data than supervised training from scratch. This additional supervised training, called fine-tuning (\ref{sec:box3}), allows a pre-trained model to learn to produce outputs for the desired downstream task \cite{Ji2021, DallaTorre2024}.
The difficulty and expense of obtaining large training datasets for many tasks makes the paradigm of pre-training and fine-tuning particularly promising.

\textbf{Pre-training tasks.}
Genomic language models process input DNA sequences into tokens representing either single nucleotides or short subsequences (akin to ``words'' in a language) using schemes such as \emph{k}-mer tokenization or byte-pair encoding.
Following approaches designed for natural language processing, genomic language models use pre-training tasks such as masked language modeling (MLM), in which a subset of tokens (often 15\% by convention) are masked or randomly mutated during pre-training, then the masked tokens are learned by the model from surrounding context.
Another pre-training task is autoregressive language modeling (ALM), in which subsequent tokens are learned from previous context. Conceptually, MLM is a ``fill-in-the blank'' approach, while ALM is predicting the next token.
Some studies suggest room for improvement in defining biologically-relevant pre-training tasks (such as incorporating reverse complement prediction \cite{Gndz2023}) rather than simply using those from natural language modeling \cite{Consens2025}.

\textbf{Architectures.} As with supervised models, language model architectures vary.
While the self-attention mechanism of the transformer architecture \cite{Vaswani2017} is central to many models, the paradigm of self-supervised pre-training is not limited to transformers, and some genomic language models have utilized different mechanisms \cite{Zaheer2020, Poli2023}.
Encoder-decoder architectures, which learn input representations via an encoder neural network and reconstructions of inputs from those representations via a decoder neural network, are common in language modeling.
Currently, most genomic language models are encoder-only models trained with MLM \cite{Consens2025}.

Several self-supervised genomic language models have been evaluated on human non-coding variant effect prediction tasks (Supplementary Table 1). We anticipate that genomic language models will need more systematic evaluation of their performance on variants, including extensive direct comparisons to supervised models, before they are widely adopted in downstream variant applications.

\vspace{2mm}
\noindent\makebox[\linewidth]{\rule{0.625\paperwidth}{0.4pt}}

\clearpage
\noindent\makebox[\linewidth]{\rule{0.625\paperwidth}{0.4pt}}
\section{Probing and fine-tuning}\label{sec:box3}
Probing and fine-tuning are two common approaches for supervised training of a pre-trained model on a new task of interest. Probing typically refers to linear probing \cite{Alain2017}, in which the pre-trained model parameters are unchanged but a new linear layer is learned from the resulting embeddings to predict task-specific data (Box figure a). Conceptually, fine-tuning involves replacing the pre-trained model outputs with outputs for a new prediction task and allowing the pre-trained model parameters to be updated while performing additional training using a task-relevant loss function and a new type of data (Box figure b). In practice, parameter-efficient fine-tuning methods that learn a smaller set of added model parameters, rather than modifying the larger set of existing ones, have become popular \cite{Hu2022LoRA, Liu2022finetuning, DallaTorre2024} (Box figure c). After fine-tuning a self-supervised genomic language model, variant effect predictions can be extracted as they are for supervised models. 
While probing and fine-tuning are particularly important concepts for genomic language models, the notion of additional supervised training on a new prediction task (e.g. a variant task) is also commonly used for supervised models \cite{Rastogi2024, Drusinsky2024, Lal2024, hingerl2025scooby}. An important advantage of probing and fine-tuning is that they typically require less data for the new prediction task than would be required to train on that task from scratch, since they leverage the biologically-relevant internal representations already captured in the pre-trained model parameters. 

\begin{figure}[h]
\centering
\includegraphics[width=0.8\linewidth]{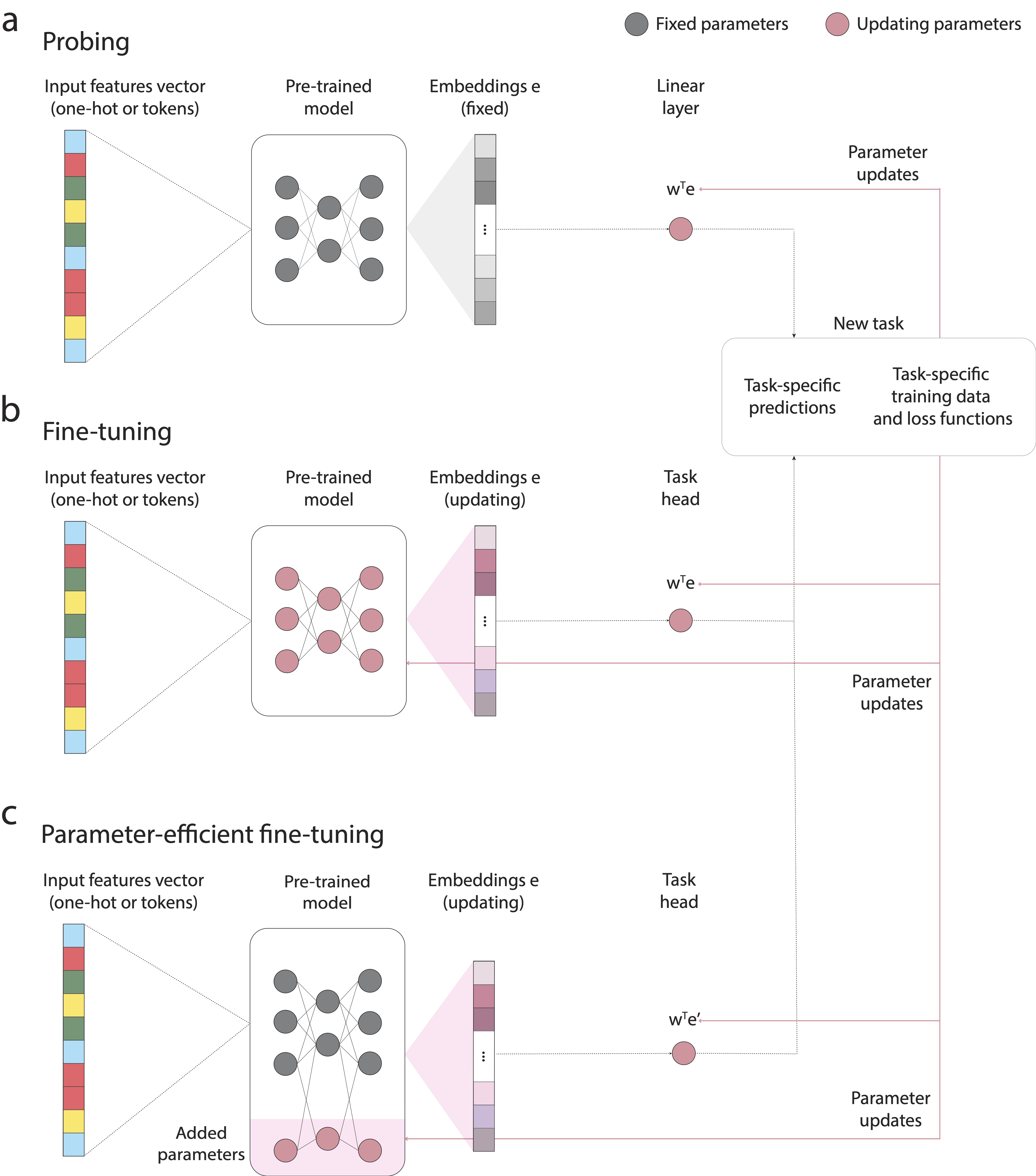}
\caption*{\textbf{Box 3 figure: Probing and fine-tuning.}}
\label{fig:probing_finetuning}
\end{figure}

\vspace{-7mm}
\noindent\makebox[\linewidth]{\rule{0.625\paperwidth}{0.4pt}}

\clearpage

\begin{figure}
\centering
\includegraphics[width=0.8\linewidth]{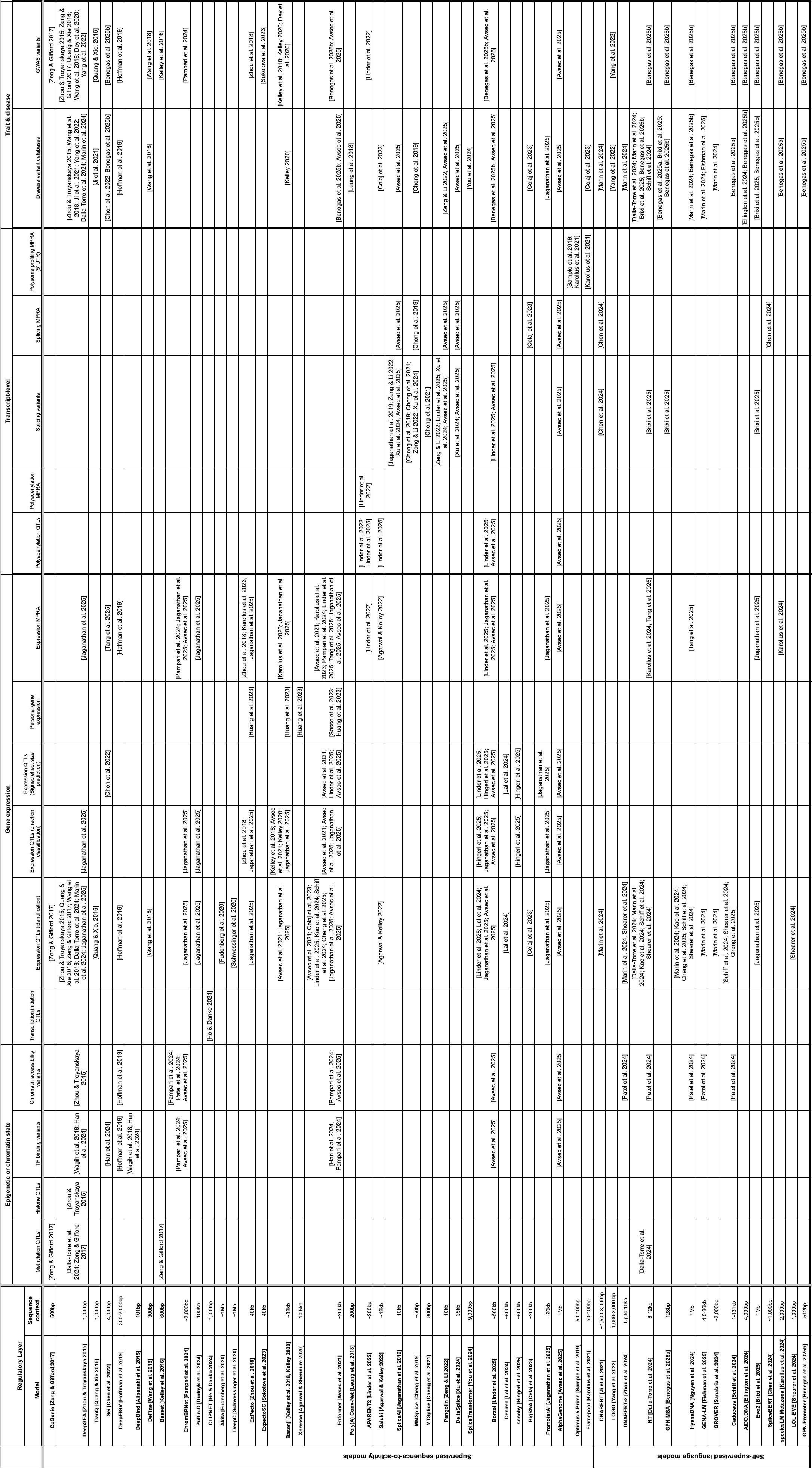}
\caption*{\textbf{Supplementary Table 1: Overview of variant effect prediction evaluations using current genomic deep learning models.} Evaluations performed either in the original modeling papers or in independent studies are indicated. Data used for evaluation are grouped into four broad categories: epigenetic or chromatin-level phenotypes, gene expression phenotypes, transcript-level phenotypes, and organismal trait or disease-related phenotypes. Models are broadly grouped by the type of data they are trained to predict, and we note that not all evaluation types may be relevant to all models. For example, epigenetic/chromatin-level evaluations are less relevant for models trained on transcript-level phenotypes (such as polyadenylation and splicing models).}
\label{fig:vep_tab1}
\end{figure}

\clearpage

\bibliography{sn-bibliography}

\end{document}